\documentclass[reprint,amsmath,amssymb,aps,prl,superscriptaddress]{revtex4-1}

\usepackage{times,color,amsmath}
\usepackage{subfigure,epsfig,graphicx,epstopdf}
\usepackage[colorlinks]{hyperref}
\hypersetup{linkcolor = {blue},	citecolor = {blue},	urlcolor = {blue}}

\newcommand{\afft}{\thanks{These authors contributed equally to this work}}
\newcommand{\affa}{\affiliation{Center for Integrated Quantum Information Technologies, School of Physics and Astronomy, State Key Laboratory of Advanced Optical Communication Systems and Networks, Shanghai Jiao Tong University, Shanghai 200240, China}}
\newcommand{\affb}{\affiliation{Synergetic Innovation Center of Quantum Information and Quantum Physics, University of Science and Technology of China, Hefei, Anhui 230026, China}}

\newcommand{\affc}{\affiliation{Guangdong Provincial Key Laboratory of Quantum Metrology and Sensing \& School of Physics and Astronomy,\\ Sun Yat-Sen University (Zhuhai Campus), Zhuhai 519082, China}}
\newcommand{\affd}{\affiliation{State Key Laboratory of Optoelectronic Materials and Technologies, Sun Yat-Sen University (Guangzhou Campus), Guangzhou 510275, China}}
\newcommand{\affe}{\affiliation{Nonlinear Physics Centre, Research School of Physics and Engineering,\\ Australian National University, Canberra ACT 2601, Australia}}

\begin{document}

\title{Constructing higher-order topological states in higher dimension}

\author{Yao Wang}		\afft \affa \affb  
\author{Yongguan Ke}   \afft \affc  \affe
\author{Yi-Jun Chang}		\affa \affb    
\author{Yong-Heng Lu}		\affa \affb      
\author{Jun Gao}				 \affa \affb		 
\author{Chaohong Lee}		\email{lichaoh2@mail.sysu.edu.cn} \affc \affd 
\author{Xian-Min Jin}		 \email{xianmin.jin@sjtu.edu.cn}  \affa \affb

\date{\today}

\maketitle

\textbf{Higher-order topological phase as a generalization of Berry phase attracts an enormous amount of research. The current theoretical models supporting higher-order topological phases, however, cannot give the connection between lower and higher-order topological phases when extending the lattice from lower to higher dimensions. Here, we theoretically propose and experimentally demonstrate a topological corner state constructed from the edge states in one dimensional lattice. The two-dimensional square lattice owns independent spatial modulation of coupling in each direction, and the combination of edge states in each direction come up to the higher-order topological corner state in two-dimensional lattice, revealing the connection of topological phase in lower and higher dimensional lattices. Moreover, the topological corner states in two-dimensional lattice can also be viewed as the dimension-reduction from a four-dimensional topological phase characterized by vector Chern number, considering two modulation phases as synthetic dimensions in Aubry-Andr\'e-Harper model discussed as example here. Our work deeps the understanding to topological phases breaking through the lattice dimension, and provides a promising tool constructing higher topological phases in higher dimensional structures.}\\

\section{Introduction}
The higher-order topological phases introduced in higher-dimensional lattices recently extend the conventional understanding on the topological nontrivial materials, where the $d$-dimensional lattice owns not only the first-order ($d-1$)-dimensional edge states but also the $n$-order ($d-n$)-dimensional edge states~\cite{T20171, T20172, T20181, T20190, T20191, T20192, T20201, T20202}. The second-order corner states in two-dimensional (2D) lattices are widely investigated since 2019 in sonic~\cite{Es1, Es2, Es3, Es4, Es5, Es6}, ring resonator~\cite{Er}, waveguide~\cite{Ew1, Ew2, Ew3}, cavity~\cite{Ec1, Ec2}, and cold atom~\cite{Ecold} systems. Recently, the higher-order topological states in three-dimensional lattices are also reported~\cite{E3D1, E3D2}. The investigations on higher-order topological phases in both theories and experiments promote and extend the development of topological photonics~\cite{Topo_review_1, Topo_review_2}.

The current principles to seek high-order topological states are mainly based on analyzing spatial or (and) nonspatial symmetries~\cite{T20171, T20172,TPRL118,T20181,Langbehn2017,Song2017,Linhu2018,Max2018}.
In spatial-symmetric (such as inversion- or rotational-symmetric) systems, high-order topological states may originate from quantized dipole polarization~\cite{TPRL118,T20181} or multipole moments~\cite{Langbehn2017,Song2017}. 
In nonspatial-symmetric (such as chiral-symmetric) systems, corner states may arise due to nontrivial edge winding numbers~\cite{{Linhu2018}}.
By combining  nonspatial and spatial symmetries, second-order topological insulator and superconductors have been partially classified~\cite{Max2018}. 
The existing schemes requiring delicate designs of overall symmetry, as a top-to-bottom approach, cannot provide insight of the connection between lower-order and higher-order topological states. 
Since lower-order edge states are well-known, we may wonder whether it is possible to use lower-order topological states as building blocks assembling to higher-order topological states.
If possible, what are their topological correspondence to the bulk?

Here, we theoretically propose and experimentally demonstrate a bottom-to-top scheme for constructing topological corner states by using topological edge states as building blocks.
In each direction the topological edge states are snapshot states in a topological pumping by means of a changing one-dimensional dipole moment which is related to Chern number.
Such scheme naturally extends Chern number to vector Chern number with individual components separately defined in each direction,  from lower- to higher-dimensional lattices.
The hierarchical relation between two-dimensional Zak phase~\cite{TPRL118,T20181} and vector Chern number can be understood as a varying of two-dimensional dipole polarization generates quantized charge pumping in two directions.
The fact that corner states are guaranteed by nontrivial vector Chern number can be termed as \emph{bulk-corner correspondence}, and they inherit the topological origin of edge states as a dimension reduction of quantum Hall phase. 
We have to emphasize that such corner states do not require any fine tuning of spatial or nonspatial symmetries. 

Taking the off-diagonal Aubry-Andr\'e-Harper (AAH) model for example, we theoretically analyze the topological origin when extending the lattice from one dimension to two dimensions. We construct the two-dimensional photonic topological lattice and successfully observe the higher-order topological corner states predicted in theory. Our model gives an intuitive understanding on the raising of higher-order topological phases in higher dimensional lattice, which connects the topological phases in different dimensional lattices and provides a convenient tool for constructing higher-order topological phases in higher dimensional lattices.

\begin{figure}
	\centering
	\includegraphics[width=1.0\linewidth]{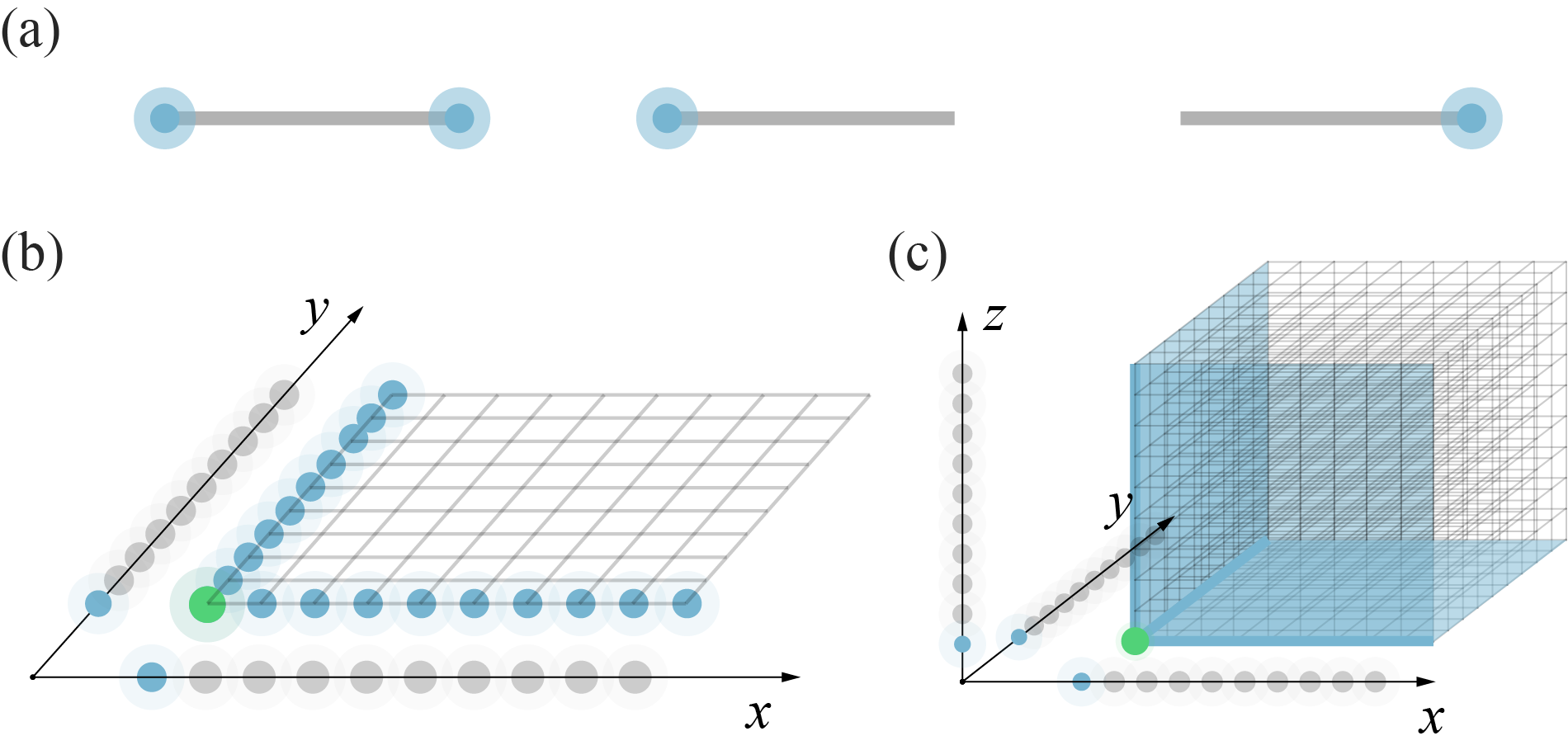}
	\caption{\textbf{Schematic of constructing corner states.} 
		\textbf{(a)} Three types of edge states in one-dimensional lattices. The edge states in one-dimensional lattice can be regarded as the building block for the higher-order topological states. 
		\textbf{(b-c)} Corner states in two- and three-dimensional lattices built by edge states in one-dimensional lattices. We can find the connection between the topological states in different dimensional lattice by projecting the higher-dimensional lattice into one-dimensional lattice in direction of x, y and z axis respectively.}
	\label{f1}
\end{figure}

\section{Vector Chern number and corner states}
We explore a systematical method to construct corner states in a two-dimensional square lattice.
Consider that the position of a particle is denoted by $(i,j)$, where $i$ and $j$ are the site indices of $x$ and $y$ directions respectively.
The coupling between $(i,j)$th and $(k,l)$th lattice sites has the form,
$H_x(i,k)\delta_{j,l}+\delta_{i,k}H_y(j,l)$, where $H_{x}(i,k)$ is the coupling matrix along $x$ direction irrelevant to positions in $y$ direction, and vice versa.
The motion of a particle hopping in such lattice is governed by the Hamiltonian,
\begin{equation}
	H=H_x\otimes I_y+I_x\otimes H_y,
\end{equation}
where $I_{s}$ is an identical matrix in $s$ direction with $s\in x, \ y$.
Once we obtain the eigenvalues $E_p^{s}$ and eigenstates $|\psi_p^{s}\rangle$ corresponding to $H_s$ where $p$ are quantum numbers,
we can immediately prove that
\begin{equation}
	H|\psi_m^{x}\rangle\otimes |\psi_n^{y}\rangle=(E_m^{x}+E_n^{y})|\psi_m^{x}\rangle\otimes |\psi_n^{y}\rangle,
\end{equation}
that is, $E_m^{x}+E_n^{y}$ and $|\psi_m^{x}\rangle\otimes|\psi_n^{y}\rangle$ are the eigenvalues and eigenstates corresponding to $H$, respectively.
If $|\psi_m^{x}\rangle$ and $|\psi_n^{y}\rangle$ are topological edge states, the product of these edge states becomes a topological corner state in two dimensions.
Hence the seeking for topological corner states is transformed to separately design the coupling matrix in each direction that supports topological edge states.

Consider that the coupling matrix $H_s$ is controlled by two parameters $\phi^{s}$ and $\theta^{s}$, which satisfied $H_s(\phi^{s},\theta^{s})=H_s(\phi^{s}+2\pi,\theta^{s}+2\pi)$.
In practice,  $\phi^{s}$ is a modulated phase, and $\theta^{s}$ is a twisted phase when a particle hopping across the boundary in $s$ direction by imposing twisted boundary condition.
To characterize the bulk topology, we can define vector Chern number $(C_{\mathbf{u}}^{x}; C_{\mathbf{v}}^{y})$ with individual components given by
\begin{equation}
	C_{\mathbf{w}}^{s}= \frac{1}{{2\pi i}}\int_{BZ} d \theta^{s}{d\phi^{s} \det [\mathcal{F}({\phi^{s},\theta^{s}})]}.
\end{equation}
Here,  $ [\mathcal{F}({\phi^{s},\theta^{s}})]^{m,n}=\partial_{\phi^{s}} A_{\theta^{s}}^{m,n}-\partial_{\theta^{s}} A_{\phi^{s}}^{m,n} +i[A_{\phi^{s}},A_{\theta^{s}}]^{m,n} $ are elements of the non-Abelian Berry curvature with elements of Berry connection  $(A_\mu)^{m,n}=\langle \psi_{m}^s({\phi^{s},\theta^{s}})|\nabla_\mu|\psi_{n}^s({\phi^{s},\theta^{s}})\rangle$. $m,n\in \mathbf{w}$ which is a  subset of near degenerate bands.
If both the two components of the vector Chern number $(C_{\mathbf{u}}^{x} ; C_{\mathbf{v}}^{y})$ are integer, in the open boundary condition there exists at least one topological corner state at some fixed modulated phases $(\phi^x,\phi^y)$.
We term this relation as \emph{bulk-corner correspondence}, which gives  a clear topological origin of corner states, i.e., a dimensional reduction of a four-dimensional topological space characterized by vector Chern number.

\begin{figure}[!t]
	\centering
	\includegraphics[width=1.0\linewidth]{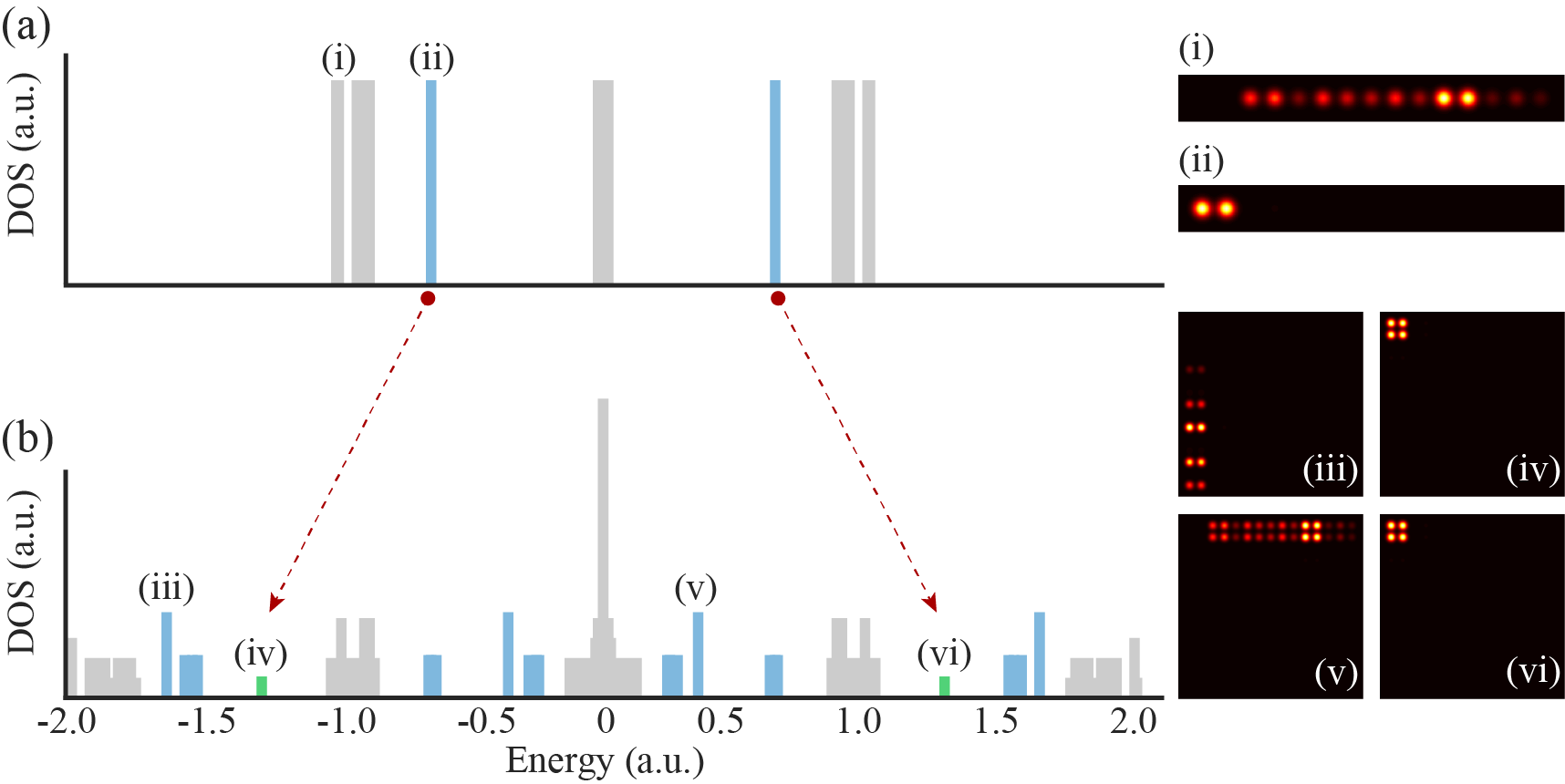}
	\caption{\textbf{The density of states for (a) one-dimensional AAH lattice and (b) two-dimensional AAH lattices.} The finite energy gaps between corner states and extended states inherit from those between edge states and extended states. The corner states are constructed from the edge states, the former has twice the energy of the latter. The local density of states are shown in the insets, the bulk state and edge state in one-dimensional AAH lattice are shown in (i) and (ii) respectively, the edge states and corner states in two-dimensional AAH lattice are shown in (iii, v) and (iv, vi) respectively. The parameters in simulation are adopted as: $t_{x(y)}=0.5$, $\lambda_{x(y)}=0.95$, $b_{x(y)}=(\sqrt{5}+1)/2$, and there are 15 sites for one-dimensional lattice and 15$\times$15 sites for two-dimensional lattice.
	DOS: density of states.}
	\label{f2}
\end{figure}

\begin{figure*}[!t]
	\centering
	\includegraphics[width=0.9\linewidth]{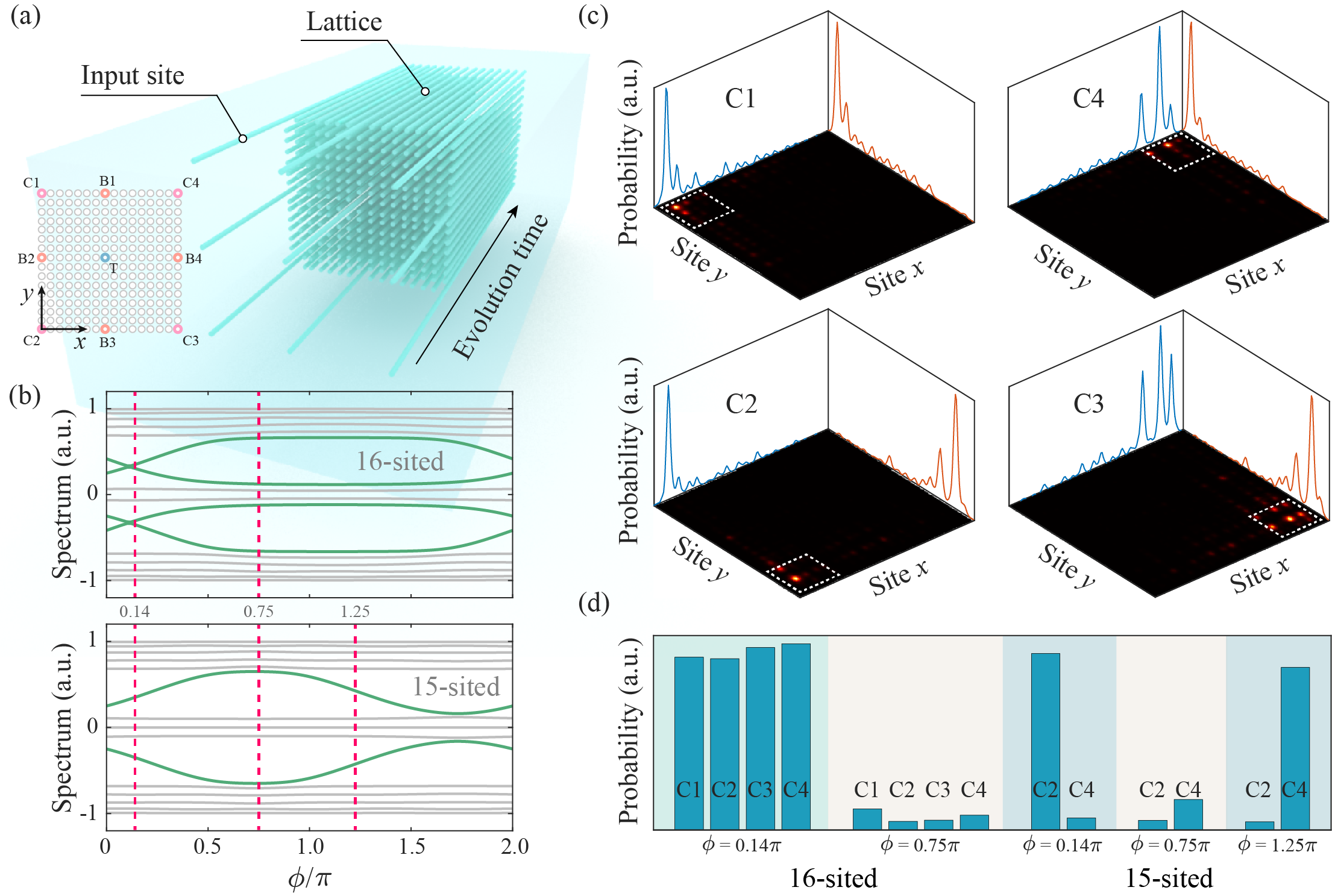}
	\caption{\textbf{Corner states.} 
		\textbf{(a)} Schematic of the fabricated photonic quasicrystal. Nine input sites are set in the lattice, four waveguides marked as C1, C2, C3 and C4 are designed to observe the corner states. Four waveguides marked as B1, B2, B3 and B4 are designed to observe the edge states. The waveguide marked as T is designed to observe bulk state.
		\textbf{(b)} Spectrum of one-dimensional \textit{off-diagonal} AAH lattice. For 16-sited lattice, two boundary modes (green lines) cross in the band gap, for 15-sited lattice, only one boundary mode connects the bands separated by the gap. The red dash lines give the $\phi$ adopted in experiment.
		\textbf{(c-d)} Measured corner states. The photons are confined in the corners when we excite the lattice corner sites (c). The white dotted squares point out the ranges of corner states. The quantified results confirm the theoretically predicted topological corner states arising by extending topological lattice with edge state from one dimension to two dimensions.}
	\label{f3}
\end{figure*}

\begin{figure*}[!t]
	\centering
	\includegraphics[width=1.0\linewidth]{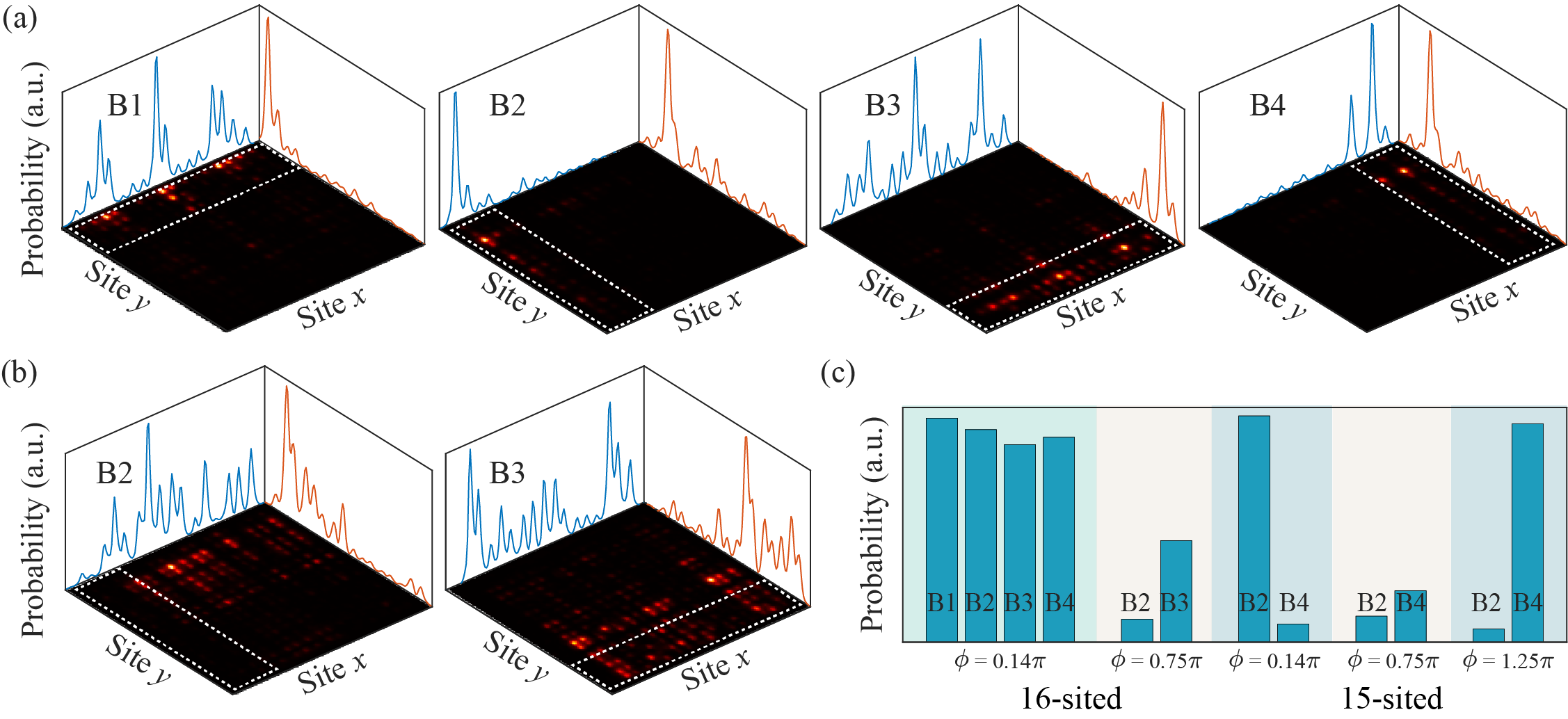}
	\caption{\textbf{Edge states.}  
		\textbf{(a)} The measured result of edge states. We inject the photons into the lattices from the input sites B1, B2, B3, and B4 respectively, the photons are localized in the edges of the lattices.
		\textbf{(b)} The trivial cases. The photon cannot be confined in the edges of the lattices. The white dotted squares point out the ranges of edge states. 
		\textbf{(c)} The quantified results. The edge states can be conveniently extended from lower dimension to higher dimension.}
	\label{f4}
\end{figure*}

For explicitness, we focus on a two-dimensional \textit{off-diagonal} AAH lattice with hopping strengths varying in space,
\begin{align}
\label{eq1}\nonumber
H = &\sum_{i,j} t_x[1+\lambda_x \cos(2\pi b_x i+\phi^{x})]\hat{a}_{i,j}\hat{a}_{i+1, j}^\dagger \\
&+t_y[1+\lambda_y \cos(2\pi b_y j+\phi^{y})]\hat{a}_{i, j}\hat{a}_{i, j+1}^\dagger + H.c.,
\end{align}
where $\hat{a}_{i, j}^\dagger$ ($\hat{a}_{i, j}$) is the creation (annihilation) operator at site ($i, j$),
$t_{x(y)}$ is the average coupling strengths, $\lambda_{x(y)}$, $b_{x(y)}$, and $\phi^{x(y)}$ are modulated strengths, frequencies and phases, respectively. 
In numerical calculations of vector Chern number, we choose $t_{x(y)}=0.5$, $\lambda_{x(y)}=0.95$, $b_{x(y)}=(\sqrt{5}+1)/2$, and the total sites are $15\times 15$ for two-dimensional lattice. 
There are three subsets of bands in each direction, and the vector Chern number takes values of $(1,-2,1;1,-2,1)$, indicating the existence of topological corner states.

In each direction, there are three types of one-dimensional topological edge states, one localized at both edges, one localized at the left edge and the last one localized at the right edge; see Fig.~\ref{f1}(a).
These edge states constitute basic building blocks to construct topological corner states in higher dimension.
As shown in Fig.~\ref{f1}(b), we can construct corner states  by using the edge states in both $x$ and $y$ directions (see Supplemental Materials for more details~\cite{SM}).
Since the couplings along $x$ and $y$ directions are independent, the robustness of corner states inherits that of one-dimensional edge states in each dimension.
Taking edge states in $x$ direction for example, there are energy gaps between edge states and extended states; see Fig.~\ref{f2}(a). %
These topological edge states are robust to considerable disorder, perturbation and long-range couplings that mix the two dimensions provided energy gaps keep open.
%
Hence, the constructing corner states also share similar topological protection to the one dimensional edge states, where there are finite energy gaps between corner states and extended states; see Fig.~\ref{f2}(b). 
Apart from corner states, products of edge state in one direction and extended state in other direction form a continuum subset which is also separated from extended states.
Such approach can be naturally generalized to construct corner in three dimensional periodically-modulated lattice, see Fig.~\ref{f1} (c), along with hinge and surface states (see Supplemental Materials for more details~\cite{SM}).

What's more, when $b_x=b_y=1/2$, the above model reduces to a two-dimensional Su-Schrieffer-Heeger (SSH) lattice, where coupling matrices are changed in a stagger way in both $x$ and $y$ directions~\cite{TPRL118}.
Indeed, topological corner states are  predicted with an alternative theory of two-dimensional polarization~\cite{T20201,T20181} and observed in experiment of photonic crystal slabs~\cite{Es1,Ew2,Ew3}.
The varying of two-dimensional polarization~\cite{TPRL118} could give rise to topologial charge pumping in two dimensions which is characterized by vector Chern number.
However, on one hand, these corner states can be more easily and naturally understood in our theoretical framework, that is, they are the product of two edge states in both $x$ and $y$ directions.  
On the other hand, in contrast to two-dimensional polarization which relies on spatial symmetries, our theory of vector Chern number can also predict corner states without requiring any fine-tuning of symmetries.

\section{Experimental implement}
In experiment, we first realize the Hamiltonian~\eqref{eq1} in a two-dimensional array of waveguides with modulated spacing. The site number is 15 or 16 in both $x$ and $y$ directions, the average coupling strength $t_x=t_y=t$ is adopted as 0.3 for photon with wavelength of 810 nm, the modulating amplitude $\lambda_x=\lambda_y=\lambda$ is set as 0.5, the periodic parameter $b_x=b_y=b$ is $(\sqrt5+1)/2$, and the initial phases $\phi^{x}$ and $\phi^{y}$ are set as the same value. We fabricate the photonic waveguide lattices according the Hamiltonian using the femtosecond laser direct-writing technique~\cite{fabri_1, fabri_2, fabri_3, fabri_4, PIT_Gap}. As shown in Fig.~\ref{f3}(a), the 
propagation direction of the waveguide lattice maps the evolution time, hence the fabricated tree dimensional waveguide lattice realizes the designed two-dimensional \textit{off-diagonal} AAH lattice. We further set nine sites for injecting the photons, including four corner sites labeled as from C1 to C4, four boundary sites labeled as from B1 to B4, and one site in the lattice center labeled as T.

According to the prediction in theory, the corner state will appear when we extend the one-dimensional topological lattice with edge state to a two-dimensional lattice, and the corresponding topological origin is also extended to higher dimensions. As shown in Fig.~\ref{f3}(b), there are edge states in both two ends of lattice for 16-sited one-dimensional \textit{off-diagonal} AAH lattice with initial phase $\phi$ = 0.14$\pi$. We fabricate the two-dimensional \textit{off-diagonal} AAH lattice with initial phase $\phi^{x}=\phi^{y}$ = 0.14$\pi$ to demonstrate the predicted corner states. We inject the photons with wavelength of 810 nm in to the lattice from four corners respectively, the photons will be confined in the excited lattice corners if there are higher-order corner states in theoretical prediction. As shown in Fig.~\ref{f3}(c), the photon output distributions after 40 mm evolution distance are localized in the white dotted squares, which give the ranges of corner states.

In Fig.~\ref{f3}(d), we give the quantified results for measured corner states, which is calculated by
\begin{equation}
	\xi_{p,q}=\sum_{i,j}\hat{a}_{i,j}^\dagger\hat{a}_{i,j} \quad (|i-p|\leq l, |j-q|\leq l),
\end{equation}
where $(p, q)$ presents the excited site indices, and $l$ describes the range of corner states adopted as 3. Compared with the 16$\times$16-sited lattice with $\phi=0.14\pi$, there is no corner state for the case of $\phi=0.75\pi$, and the photons flow out of the corner state range. This is because there is no edge state in the one-dimensional AAH lattice for the case of $\phi=0.75\pi$. Furthermore, we fabricate three two-dimensional 15$\times$15-sited lattices with phase $\phi=0.14\pi$, 0.75$\pi$, and 1.25$\pi$ respectively. 
There is only left (right) edge state for one-dimensional lattice with $\phi=0.14\pi$ (1.25$\pi$), therefore the corner state can only be observed by exciting the lattice from input C2 (C4). Similar to 16$\times$16-sited lattice, there is no corner state for the case of $\phi=$ 0.75$\pi$. We excite the lattices from input C2 and C4 respectively and measure the photon output distributions, the quantified results, together with the results of 16$\times$16-sited lattices, confirm theoretical predictions on topological corner states arising by extending topological lattice with edge state from one dimension to two dimension.

The corner states appearing in two-dimensional lattice require combination of the one-dimensional lattices owning edge states in both the $x$ and $y$ directions. Differently, The edge states in higher dimensional lattices, as a product of edge state in one direction and extended state in other direction, can be naturally extended from the edge states in lower dimensional lattices. As shown in Fig.~\ref{f4}(a), we inject the photons into the 16$\times$16-sited lattice with $\phi$ = 0.14$\pi$ from the input B1, B2, B3 and B4 respectively, the photons are confined in the boundaries. For the case of $\phi$ = 0.75$\pi$, there is no edge state that can be extended, so we can find that the photons flow out the boundary ranges, as shown in Fig.~\ref{f4}(b) taking the cases of B2 and B3 for example. The quantified results in Fig.~\ref{f4}(c) show the observed edge states extended from one-dimensional lattices. As intuitive understanding, edge states are extended from dots to lines when the topological lattices are extended from one dimension to two dimensions.

In conclusion, we present a theoretical explanation on topological origin of higher-order topological phase in higher dimensional lattices, which is connected to the topological phase in lower dimensional lattices. We experimentally observe the theoretically predicted higher-order topological corner states in two-dimensional \textit{off-diagonal} AAH lattices. Our model intuitively explains the connection of topological phases in different dimensional lattices, which is universal to various models and is promising to be a convenient and practical tool for constructing higher-order topological phases in higher-order lattices.\\

\begin{acknowledgments}
	The authors thank Yidong Chong and Jian-Wei Pan for helpful discussions.
	X.-M.J. is supported by 
	National Key R\&D Program of China (2019YFA0308700, 2017YFA0303700),
	National Natural Science Foundation of China (11761141014, 61734005, 11690033),
	Science and Technology Commission of Shanghai Municipality (17JC1400403),
	Shanghai Municipal Education Commission (2017-01-07-00-02-E00049).
	X.-M.J. acknowledges additional support from a Shanghai talent program.
	C. Lee is supported by 
	the Key-Area Research and Development Program of GuangDong Province (2019B030330001), 
	the National Natural Science Foundation of China (11874434, 11574405), 
	and the Science and Technology Program of Guangzhou (201904020024). 
	Y. K. is partially supported by the Office of China Postdoctoral Council (20180052), 
	the National Natural Science Foundation of China (11904419), 
	and the Australian Research Council (DP200101168).
	
\end{acknowledgments}

\clearpage

\renewcommand{\bibnumfmt}[1]{[S#1]}  \renewcommand{\citenumfont}[1]{S#1}
\renewcommand{\thefigure}{S\arabic{figure}}  \renewcommand{\theequation}{S\arabic{equation}}
\setcounter{equation}{0}
\setcounter{figure}{0}

\section{Supplementary Materials: \\Constructing higher-order topological states in higher dimension}
~\\
\subsection{Constructing corner states with off-diagonal AAH model}
\begin{figure*}[!htp]
	\centering
	\includegraphics[width=0.9\linewidth]{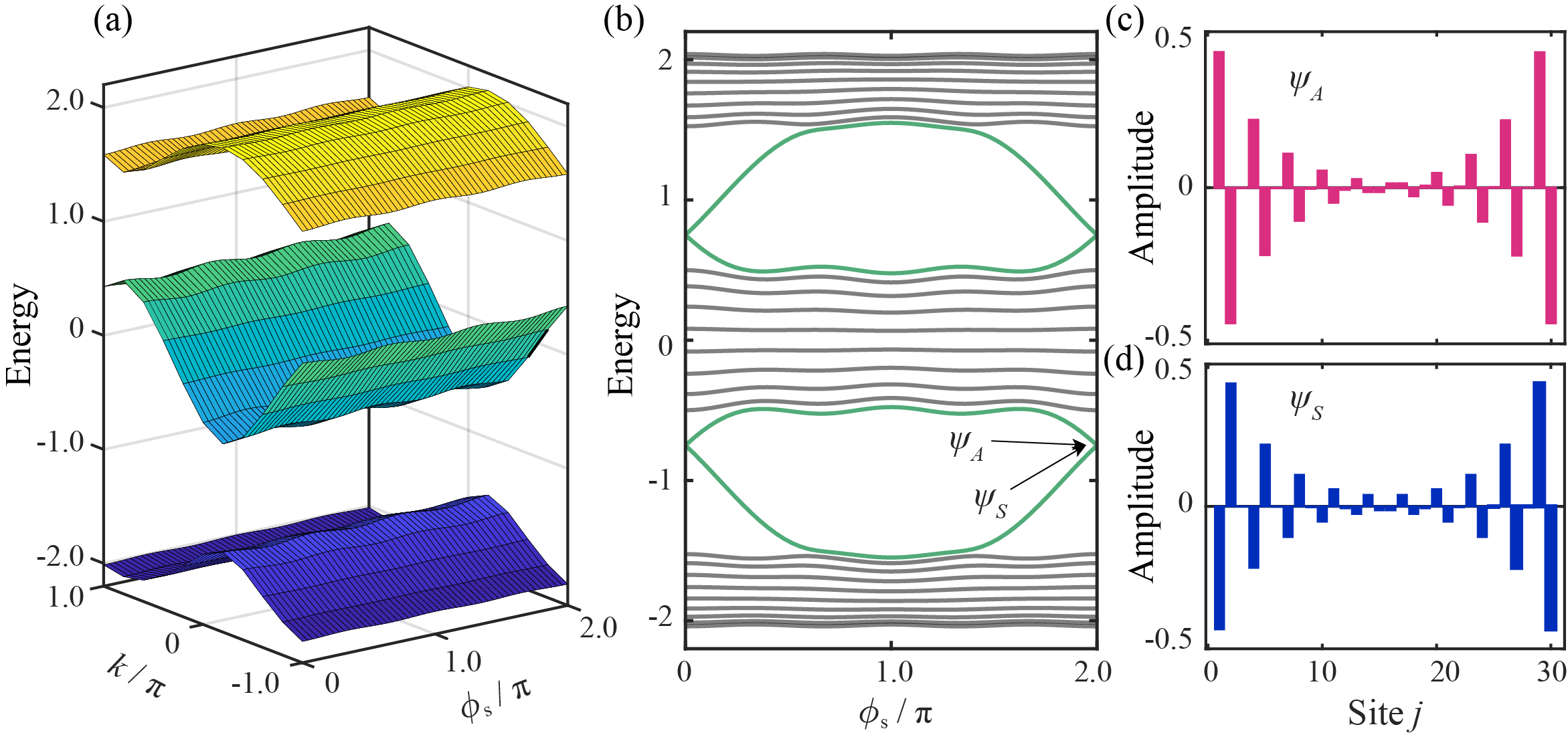}
	\caption{\textbf{Appearance of edge modes and their topological origin.} (\textbf{a}) Energy bands in ($k, \phi^{s}$) space. The corresponding Chern numbers for three bands are $(-1,2,-1)$. (\textbf{b}) Energy spectrum as a function of $\phi^{s}$ under open boundary condition.  (\textbf{c}) Asymmetry edge state and  (\textbf{d}) Symmetry edge state for $\phi^{s}=0(2\pi)$. The parameters are chosen as $t_{s}=1$, $\lambda_s=0.5$, $b_s=1/3$, and the system size is $30$. } 
	\label{Bulk_Boundary}
\end{figure*}

The Hamiltonian~(4) in the main text can be rewritten as $H=H_x\otimes I_y+I_x\otimes H_y$ with
\begin{equation}
	H_{s}=\sum_{j} \{t_{s}[1+\lambda_{s} \cos(2\pi b_{s} j+\phi_{s})]\hat{a}_{j}\hat{a}_{j+1}^\dagger+H.c.\}.  
	\label{AAH1D}
\end{equation}
Before proceeding to corner states, we first study the topological properties of the Hamiltonian~\eqref{AAH1D} with rational modulated frequency $b_s=\mu/\nu$, where $\mu$ and $\nu$ are coprime numbers.
In this case, we can calculate the vector Chern number in an alternative way.
When the twisted angle is $0$ (i.e., in periodic boundary condition),
the system is invariant when translating a particle from the $j$th to $(j+\nu)$th sites, that is, the system has translational symmetry.
According to Bloch theorem, the eigenstates are Bloch functions with quasi-momentum $k$, and the eigenvalues form $\nu$ Bloch bands.
Here, the quasi-momentum is a good quantum number and provides an intrinsic parameter to define topological invariant.  
Instead of integral over twisted angle, each component of vector Chern number can be defined in the space formed by quasi-momentum and modulated phase,
\begin{equation}
	{C_{n}^s} = \frac{1}{{2\pi i }}  \int_{BZ} dk {d\phi^{s}  \left( { \langle {\partial _{\phi^{s}} }\psi_n^s |{\partial _k}\psi_n^s \rangle } \right) -\langle {\partial _k}\psi_n^s |{\partial _{\phi^{s}} }\psi_n^s \rangle  }. 
\end{equation}
Such definition is consistent with Eq.~(3) in the main text.
According to bulk-boundary correspondence, when the sum of all the Chern numbers below a gap is non-zero, edge modes should appear in this gap.
To show this, in the periodic boundary condition we calculate the energy bands in the ($k,\phi^{s}$) parameter space and their Chern numbers, and in the open boundary condition we calculate energy spectrum as a function of modulated phase $\phi^{s}$; see Fig.~\ref{Bulk_Boundary}.    
The parameters are chosen as $t_{s}=1$, $\lambda_s=0.5$, $b_s=1/3$, and the system size is $30$.
There are three bands in each direction and the corresponding vector Chern number is $(-1,2,-1; -1,2,-1)$.
Indeed, in each direction, there are isolated edge modes in each energy gap.
When $\phi^{s}=0(2\pi)$, the two edge states in the first gap become degenerate and their wave-functions  are asymmetric [Fig.~\ref{Bulk_Boundary}(c)] and symmetric [Fig.~\ref{Bulk_Boundary}(d)]. 
Here, we have clearly shown that the appearance of the edge states results from dimension reduction of  topological phases in the two-dimensional parameter space $(k,\phi^{s})$.

\begin{figure*}[!htp]
	\centering
	\includegraphics[width=0.95\linewidth]{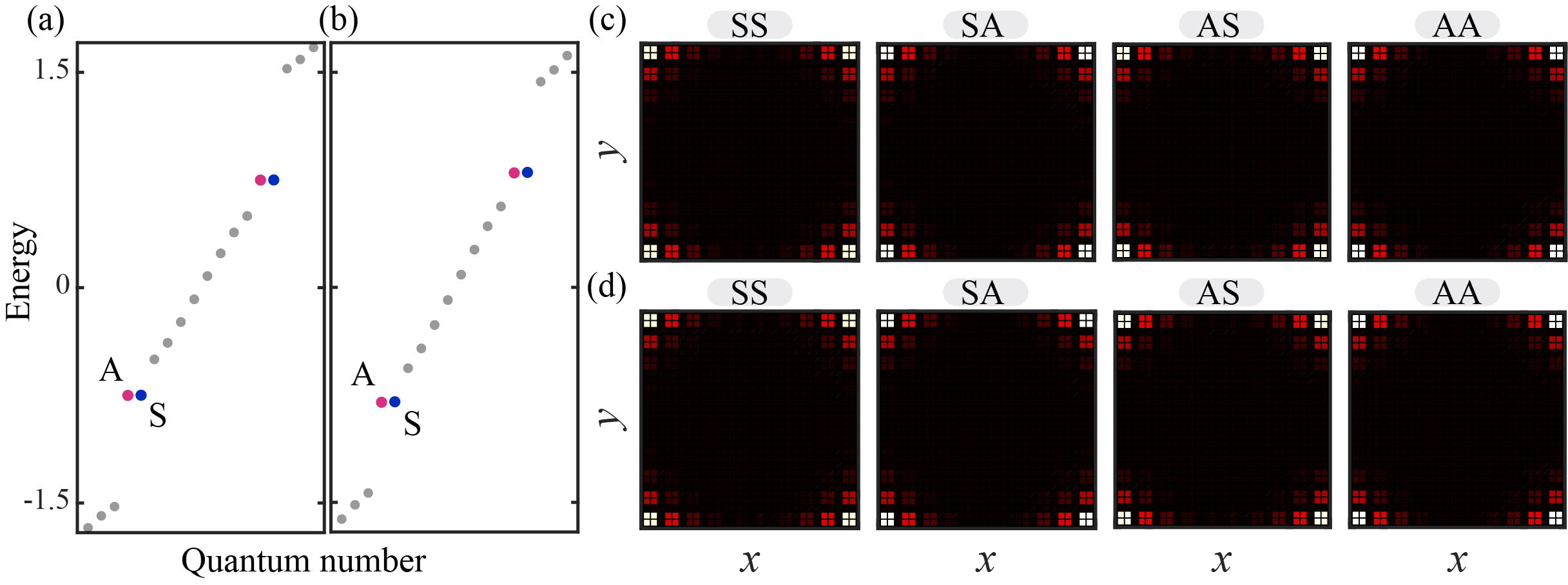}
	\caption{\textbf{Constructing corner states with edge states in $x$ and $y$ directions.}  (\textbf{a}) and (\textbf{b}): Energy spectral for off-diagonal AAH models in $y$ and $x$ directions, respectively.  (\textbf{c}) corner states obtained by combining symmetric and asymmetric edge states in $x$ and $y$ directions. (\textbf{d}) The corner states obtained by diagonalizing Hamiltonian~(4) in the main text with the same energies as those in the top panel.  The parameters are chosen as $t_{x}=t_y=1$, $\lambda_x=0.4$, $\lambda_y=0.5$, $b_x=b_y=1/3$, $\phi^x=\phi^y=0$ and the numbers of lattices in both $x$ and $y$ directions are $30$. } 
	\label{Ansatz}
\end{figure*}

With the edge states at hand, we can construct the corner states with the method depicted in the main text.
For a set of degenerate eigenstates, the superposition of these eigenstates is also the eigenstates. 
To avoid unnecessary degeneracy, we consider the parameters for off-diagonal AAH model in the $x$ direction are slightly different from those in the $y$ dimension. 
The parameters are chosen as $t_{x}=t_y=1$, $\lambda_x=0.4$, $\lambda_y=0.5$, $b_x=b_y=1/3$, $\phi^x=\phi^y=0$ and the numbers of lattices in both $y$ and $x$ directions are $30$.
The eigenvalues of the off-diagonal AAH models in $y$ and $x$ directions are shown in Fig.~\ref{Ansatz}(a) and (b), respectively.
Since the parameters are slightly different, the energy spectral in $x$ and $y$ directions are quite similar.
The symmetric and asymmetric edge modes are marked by `S' and `A' in the spectral.
With the edge modes in the first gap, we can construct four corner states (namely, $SS$, $SA$, $AS$ and $AA$) by any combination of the symmetric and asymmetric edge states in two dimensions; see Fig.~\ref{Ansatz}(c) respectively.
For comparison, we show the corner states obtained by directly diagonalizing Hamiltonian~(4) in the main text; see Fig.~\ref{Ansatz}(d).
The eigenvalues and spatial distributions of the corner states in the bottom panel are exactly the same as those in the top panel. 
It means that the constructed corner states are indeed the eigenstates of the two-dimensional off-diagonal AAH model.

For irrational modulated frequency, translational symmetry is broken and quasi-momentum is no-longer a good quantum number to define topological invariant. 
However, vector Chern number for the irrational off-diagonal AAH model can still be defined by introducing twisted angles, as shown in the main text.
The vector Chern number well characterizes the topology of such irrational off-diagonal AAH model.
Topological corner states also appear in the energy gap as a consequence of such nontrival vector Chern number.
We show experimental observation of corner states based on the irrational off-diagonal AAH model in the main text.

\subsection{Generally constructing corner states in two-dimensional lattice}
\begin{figure}[!htp]
	\centering
	\includegraphics[width=1.0\linewidth]{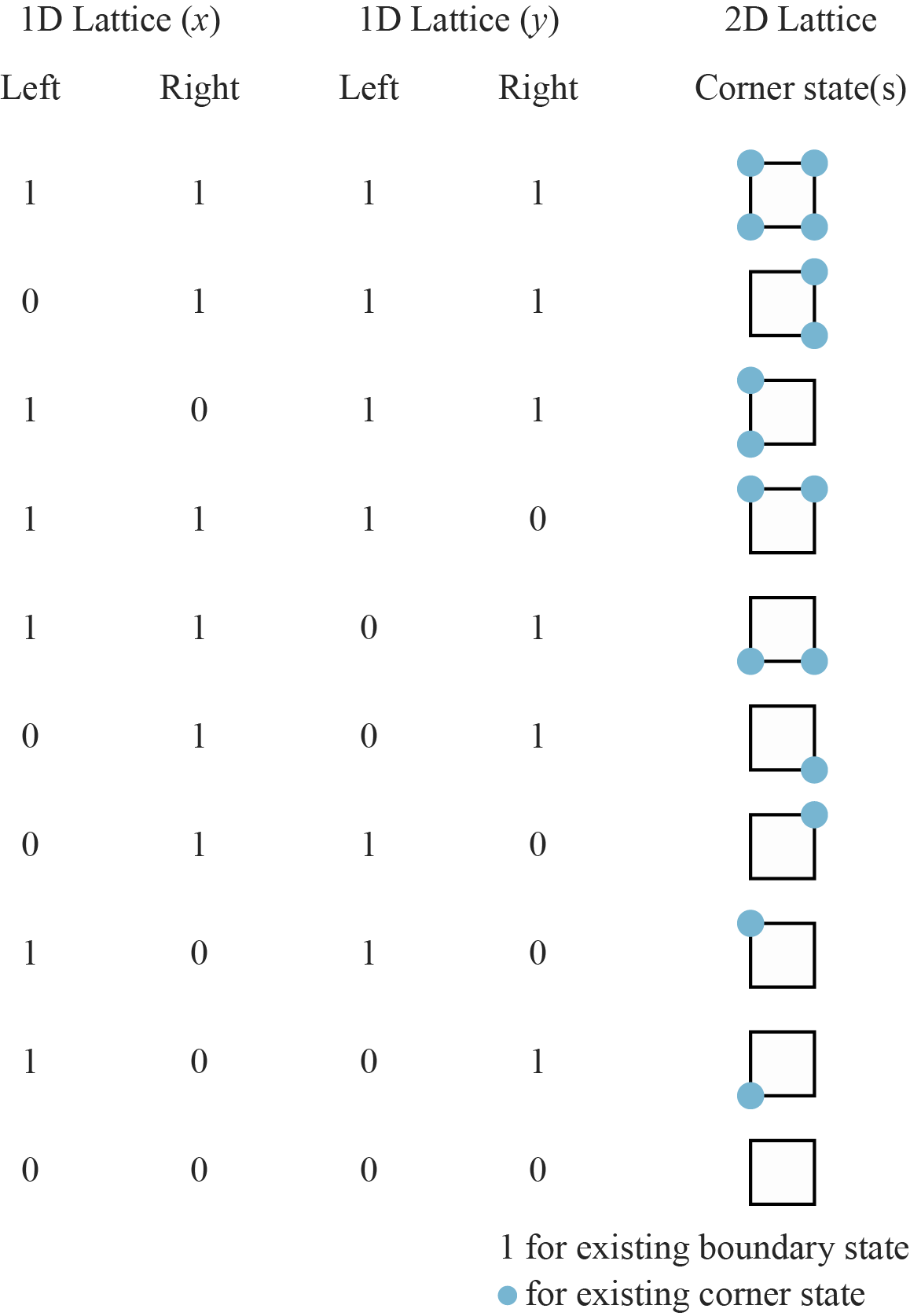}
	\caption{\textbf{Constructing corner states in two-dimensional lattice.} The second-order states can be assembled using the edge states in one-dimensional lattices.}
	\label{s_combination}
\end{figure}
As we have shown in main text and last section by taking the examples, one is able to freely construct the higher-order topological phase in two-dimensional lattice by assembling edge states in one-dimensional lattices. In this way, the higher-order states can be assembled using the edge states in lower-dimensional lattices, similar to playing LEGO game. In Fig.~\ref{s_combination}, we show the combination way to constructing corner states in two-dimensional lattice using edge states in $x$ and $y$ directional one-dimensional lattices. For example, we can choose the modulation of the lattice owning both left and right edge states in $x$ direction and just owning left edge state in $y$ direction to construct the corner states in top left and top right corners.

\subsection{Constructing corner states in three dimensional lattice}

Here we extend our approach to a three-dimensional cubic lattice model.
The position of a particle is denoted by $(i,j,k)$ which are site indices of $x$, $y$ and $z$ directions respectively.
The coupling between $(i,j,k)$th and $(l,m,n)$th lattice sites has the form,
$H_x(i,l)\delta_{j,m}\delta_{k,n}+\delta_{i,l}H_y(j,m)\delta_{k,n}+\delta_{i,l}\delta_{j,m}H_z(k,n)$, where $H_{x}(i,k)$ is the coupling matrix along $x$ direction irrelevant to positions in $y$ and $z$ directions, and vice versa. 
The motion of a particle hopping in such lattice is governed by the Hamiltonian,
\begin{equation}
H=H_x\otimes I_y\otimes I_z+I_x\otimes H_y\otimes I_z+I_x\otimes I_y\otimes H_z,
\end{equation}
where $I_{\{x,y,z\}}$ are identical matrices in $\{x, y,z\}$ directions.  
Solving the eigenproblems,
\begin{eqnarray}
H_x|\psi_p^{x}\rangle&=&E_p^{x}|\psi_p^{x}\rangle,\nonumber \\
H_y|\psi_q^{y}\rangle&=&E_q^{y}|\psi_q^{y}\rangle,\nonumber \\
H_z|\psi_l^{z}\rangle&=&E_l^{z}|\psi_l^{z}\rangle,
\end{eqnarray}
we can obtain the eigenvalues $\{E_p^{x},E_q^{y},E_l^{z}\}$ and eigenstates $\{|\psi_p^{x}\rangle,|\psi_q^{y}\rangle,|\psi_l^{z}\rangle\}$ corresponding to $\{H_x,H_y,H_z\}$.
We can immediately prove that
\begin{eqnarray}
&&H|\psi_p^{x}\rangle\otimes |\psi_q^{y}\rangle\otimes |\psi_l^{z}\rangle \nonumber \\
&=&(E_p^{x}+E_q^{y}+E_l^{z})|\psi_p^{x}\rangle\otimes |\psi_q^{y}\rangle\otimes |\psi_l^{z}\rangle, 
\end{eqnarray}
that is, $E_p^{x}+E_q^{y}+E_l^{z}$ and $|\psi_p^{x}\rangle\otimes |\psi_q^{y}\rangle\otimes |\psi_l^{z}\rangle$ are the eigenvalues and eigenstates corresponding to $H$, respectively.
If $\{|\psi_p^{x}\rangle,|\psi_q^{y}\rangle,|\psi_l^{z}\rangle\}$  are the edge states, the product of the three edge states becomes a corner state in three dimensions.

Similarly, we can also construct hinge state by selecting edge states in any two dimensions and extended state in the other dimension, and the surface states by constructing extended states in any two dimensions and edge state in the other dimension. 

 \begin{figure}[!htp]
	\centering
	\includegraphics[width=1.0\linewidth]{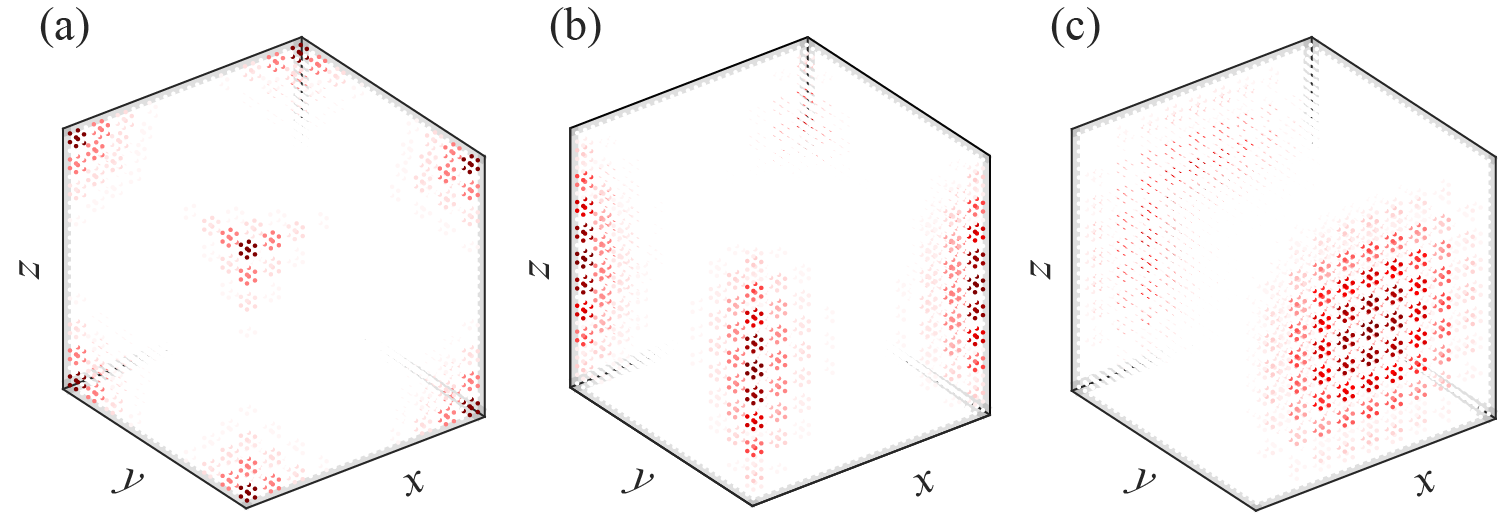}
	\caption{\textbf{Constructing (a) corner state, (b) Hinge state and (c) Surface state in a three-dimensional lattice.} The parameters are chosen as $t_{x}=t_y=t_z=1$, $\lambda_x=\lambda_y=\lambda_z=0.5$, $b_x=b_y=b_z=1/3$, $\phi^x=\phi^y=\phi_z=0$ and the numbers of lattices in $x$, $y$ and $z$ directions are $30$. }
	\label{Corner_Hinger_Surface}
\end{figure}
\begin{figure}[!htp]
	\centering
	\includegraphics[width=1.0\linewidth]{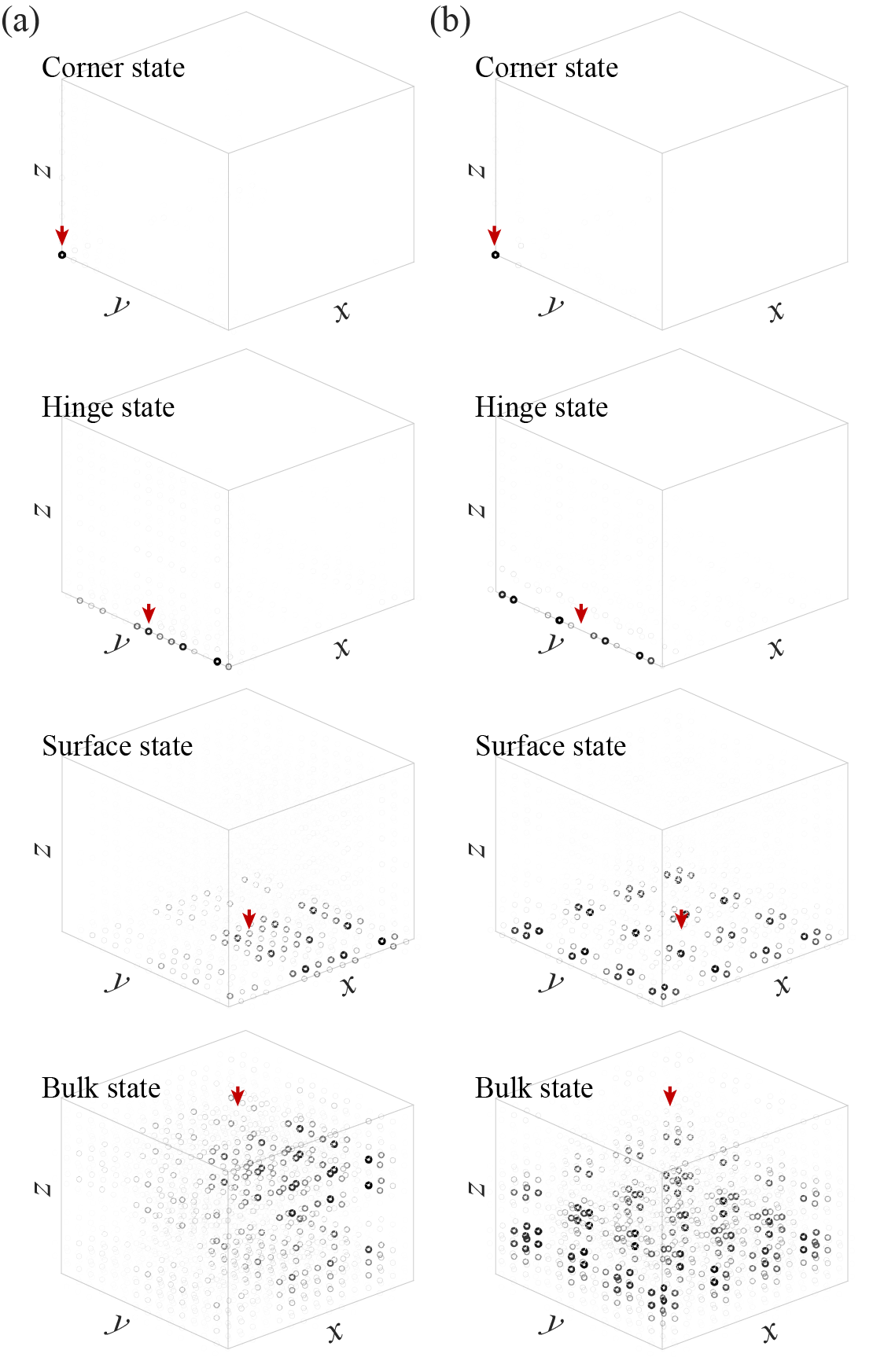}
	\caption{\textbf{Constructing higher-order topological states in three-dimensional lattice.} Higher-order topological states are observed in quasi-crystal lattice (a) and SSH lattice (b) in simulation. The line width and color depth of markers show the output probabilities. The red arrows point out the exciting position in simulation.}
	\label{s_3D lattice}
\end{figure}

In Fig.~\ref{Corner_Hinger_Surface}, we show the construction of corner, hinge and surface states in a three-dimensional lattice with $H_s$ given by Eq.~\eqref{AAH1D}.  
The parameters are chosen as $t_{x}=t_y=t_z=1$, $\lambda_x=\lambda_y=\lambda_z=0.5$, $b_x=b_y=b_z=1/3$, $\phi^x=\phi^y=\phi_z=0$ and the numbers of lattices in $x$, $y$ and $z$ directions are $30$.
The corner state is the product of the first symmetric edge states [as shown in Fig.~\ref{Ansatz}(d)] in $x$, $y$ and $z$ directions.
The Hinge state is the product of the  extended state with the highest energy of the first band in $z$ direction and the first symmetric edge states in $x$ and $y$ directions.
The Surface state is the product of the extended states with the highest energy of the first band in $x$ and $z$ directions and the first symmetric edge state in $y$ direction.

To simulate the observation of higher-order topological states in three-dimensional lattices, 
in the Fig.~\ref{s_3D lattice}, we show the evolution results of photons in assembled higher-order topological three-dimensional lattices.
In the first example, the parameters are chosen as $t_{x(y,z)}=0.5$, $\lambda_{x(y,z)}=0.5$, $\phi_{x(y,z)}=0.15$, $b_{x(y,z)}=(\sqrt{5}+1)/2$, the site numbers of lattice in both $x$ and $y$ directions are $16$, and the site number of lattice in $z$ direction is $15$. This is a three-dimensional quasi-crystal lattice based on AAH model, and the simulated results is shown in Fig.~\ref{s_3D lattice}(a).
In the second example, the parameters are chosen as $t_{x(y,z)}=0.5$, $\lambda_{x(y,z)}=0.5$, $\phi_{x(y,z)}=0$, $b_{x(y,z)}=1/2$, the site numbers of lattice in both $x$ and $y$ directions are $16$, and the site number of lattice in $z$ direction is $15$. This is a three-dimensional SSH lattice, and the simulated results is shown in Fig.~\ref{s_3D lattice}(b).


\begin{thebibliography}{99}
	\bibitem{T20171} Benalcazar, W. A., Bernevig, B. A., and Hughes, T. L., Quantized electric multipole insulators, Science \textbf{357}, 61 (2017).
	\bibitem{T20172} Benalcazar, W. A., Bernevig, B. A., and Hughes, T. L., Electric multipole moments, topological multipole moment pumping, and chiral hinge states in crystalline insulators, Phys. Rev. B \textbf{96}, 245115 (2017).
	
	\bibitem{T20181} Xie, B.-Y. \textit{et al.} Second-order photonic topological insulator with corner states, Phys. Rev. B \textbf{98}, 205147 (2018).
	\bibitem{T20190} Benalcazar, W. A., Li, T., and Hughes, T. L., Quantization of fractional corner charge in ${C}_{n}$-symmetric higher-order topological crystalline insulators, Phys. Rev. B \textbf{99}, 245151 (2019).
	\bibitem{T20191} Chen, Z.-G., Xu, C., Al Jahdali, R., Mei, J., and Wu, Y., Corner states in a second-order acoustic topological insulator as bound states in the continuum, Phys. Rev. B \textbf{100}, 075120 (2019).
	\bibitem{T20192} Luo, X.-W. and Zhang, C., Higher-Order Topological Corner States Induced by Gain and Loss, Phys. Rev. Lett. \textbf{123}, 073601 (2019).
	\bibitem{T20201} Benalcazar, W. A. and Cerjan, A., Bound states in the continuum of higher-order topological insulators, Phys. Rev. B \textbf{101}, 161116 (2020).
	\bibitem{T20202} Petrides, I. and Zilberberg, O., Higher-order topological insulators, topological pumps and the quantum Hall effect in high dimensions, Phys. Rev. Research \textbf{2}, 022049 (2020).
	
	
	\bibitem{Es1} Chen, X.-D., Deng, W.-M., Shi, F.-L., Zhao, F.-L., Chen, M., and Dong, J.-W., Direct Observation of Corner States in Second-Order Topological Photonic Crystal Slabs, Phys. Rev. Lett. \textbf{122}, 233902 (2019).
	\bibitem{Es2} Ni, X., Weiner, M., Alù, A., and Khanikaev, A. B., Observation of higher-order topological acoustic states protected by generalized chiral symmetry, Nat. Mater. \textbf{18}, 113 (2019).
	\bibitem{Es3} Xie, B.-Y. \textit{et al.} Visualization of Higher-Order Topological Insulating Phases in Two-Dimensional Dielectric Photonic Crystals, Phys. Rev. Lett. \textbf{122}, 233903 (2019).
	\bibitem{Es4} Xue, H., Yang, Y., Gao, F., Chong, Y., and Zhang, B., Acoustic higher-order topological insulator on a kagome lattice, Nat. Mater. \textbf{18}, 108 (2019).
	\bibitem{Es5} Zhang, X. \textit{et al.} Second-order topology and multidimensional topological transitions in sonic crystals, Nat. Phys. \textbf{15}, 582 (2019).
	\bibitem{Es6} Li, M. \textit{et al.} Higher-order topological states in photonic kagome crystals with long-range interactions, Nat. Photon. 14, \textbf{89} (2020).
	\bibitem{Er} Mittal, S. \textit{et al.} Photonic quadrupole topological phases, Nat. Photon. \textbf{13}, 692 (2019).
	\bibitem{Ew1} El Hassan, A., Kunst, F. K., Moritz, A., Andler, G., Bergholtz, E. J., and Bourennane, M., Corner states of light in photonic waveguides, Nat. Photon. \textbf{13}, 697 (2019).
	\bibitem{Ew2} Cerjan,A.,  J{\"u}rgensen, M., Benalcazar, W. A., Mukherjee, S., and Rechtsman, M. C., Observation of a higher-order topological bound state in the continuum, \textit{preprint in arXiv:2006.06524} (2020).
	\bibitem{Ew3} Wang, Y. \textit{et al.} Protecting Quantum Superposition and Entanglement with Photonic Higher-Order Topological Crystalline Insulator, \textit{preprint in arXiv:2006.07963} (2020).
	
	\bibitem{Ec1} Ota, Y. \textit{et al.} Photonic crystal nanocavity based on a topological corner state, Optica \textbf{6}, 786 (2019).
	\bibitem{Ec2} Dutt, A., Minkov, M., Williamson, I. A. D., and Fan, S., Higher-order topological insulators in synthetic dimensions, Light-Sci. Appl. \textbf{9}, 131 (2020).
	\bibitem{Ecold} Kempkes, S. N. \textit{et al.} Robust zero-energy modes in an electronic higher-order topological insulator, Nat. Mater. \textbf{18}, 1292 (2019).
	\bibitem{E3D1} Zhang, X. \textit{et al.} Dimensional hierarchy of higher-order topology in three-dimensional sonic crystals, Nat. Comm. \textbf{10}, 5331 (2019).
	\bibitem{E3D2} Liu, S. \textit{et al.} Octupole corner state in a three-dimensional topological circuit, Light-Sci. Appl. \textbf{9}, 145 (2020).
	
	\bibitem{Topo_review_1} Lu, L., Joannopoulos, J. D., and Solja{\v{c}}i{\'c}, M., Topological Photonics, Nat. Photon. \textbf{8}, 821 (2014).
	\bibitem{Topo_review_2} Ozawa, T. \textit{et al.} Topological photonics. Rev. Mod. Phys. \textbf{91}, 015006 (2019).
	
	\bibitem{TPRL118} Liu F. and Wakabayashi, K., Novel Topological Phase with a Zero Berry Curvature, Phys. Rev. Lett. \textbf{118}, 076803 (2017).
	
	\bibitem{Langbehn2017} J. Langbehn, Y. Peng, L. Trifunovic, F. v. Oppen, and P. W. Brouwer, Reflection-Symmetric Second-Order Topological Insulators and Superconductors. Phys. Rev. Lett. \textbf{119}, 246401 (2017).
	
     \bibitem{Song2017} Z. Song, Z. Fang, and C. Fang, (d-2)-Dimensional Edge States of Rotation Symmetry Protected Topological States. Phys. Rev. Lett. \textbf{119}, 246402 (2017).
	
	\bibitem{Linhu2018} L. Li, M. Umer, and J. Gong, Direct prediction of corner state configurations from edge winding numbers in two- and three-dimensional chiral-symmetric lattice systems, Phys. Rev. B \textbf{98}, 205422 (2018).
	
	\bibitem{Max2018} M. Geier, L. Trifunovic, M. Hoskam, and P. W. Brouwer, Second-order topological insulators and superconductors with an order-two crystalline symmetry, Phys. Rev. B \textbf{97}, 205135 (2018).
	
	
	\bibitem{SM} More detailed discussion can be found in Supplemental Materials.
	
	\bibitem{AAHPRL2012} Lang, L.-J., Cai, X., and Chen, S., Edge States and Topological Phases in One-Dimensional Optical Superlattices, Phys. Rev. Lett. \textbf{108}, 220401 (2012).
	\bibitem{AAHPRL2012p} Kraus, Y. E., Lahini, Y., Ringel, Z., Verbin, M., and Zilberberg, O., Topological States and Adiabatic Pumping in Quasicrystals, Phys. Rev. Lett. \textbf{109}, 106402 (2012).
	\bibitem{AALPRA2016} Li, J.-X., Xu, Y., Dai, Q.-F., Lan, S., and Tie, S.-L., Manipulating light–matter interaction in a gold nanorod assembly by plasmonic coupling, Laser Photonics Rev. \textbf{10}, 826 (2016).
	\bibitem{AAH20191} Wang, Y. \textit{et al.} Quantum Topological Boundary States in Quasi-Crystals, Adv. Mater. \textbf{31}, 1905624 (2019).
	\bibitem{AAH20192} Wang, Y. \textit{et al.} Topological protection of two-photon quantum correlation on a photonic chip, Optica \textbf{6}, 955 (2019).
	\bibitem{AAHPRA2019} Martinez Alvarez, V. M. and Coutinho-Filho, M. D., Edge states in trimer lattices, Phys. Rev. A 99, 013833 (2019).
	
	\bibitem{fabri_1} Davis, K. M., Miura, K., Sugimoto, N. \& Hirao, K. Writing waveguides in glass with a femtosecond laser. Opt. Lett. \textbf{21}, 1729 (1996).
	\bibitem{fabri_2} Szameit, A., Dreisow, F., Pertsch, T., Nolte, S. \& T{\"u}nnermann, A. Control of directional evanescent coupling in fs laser written waveguides. Opt. Express \textbf{15}, 1579 (2007).
	\bibitem{fabri_3} Crespi, A. \textit{et al.} Integrated multimode interferometers with arbitrary designs for photonic boson sampling. Nat. Photon. \textbf{7}, 545 (2013).
	\bibitem{fabri_4} Chaboyer, Z., Meany, T., Helt, L. G., Withford, M. J. \& Steel, M. J. Tunable quantum interference in a 3D integrated circuit. Sci. Rep. \textbf{5}, 9601 (2015).	
	\bibitem{PIT_Gap} Wang, Y. \textit{et al.} Parity-Induced Thermalization Gap in Disordered Ring Lattices. Phys. Rev. Lett. \textbf{122}, 013903 (2019).
	
	
\end{thebibliography}
\end{document}